\documentclass[traditabstract]{aa} 

\pdfoutput=1
\usepackage{amsmath}
\usepackage{graphicx}
\usepackage{txfonts}
\usepackage{natbib}

\def\pccp{PCCP}
\def\jpca{J. Phys. Chem. A}
\def\jpc{J. Phys. Chem.}
\def\cp{Chem. Phys.}
\def\jms{J. Mol. Spectrosc.}
\bibpunct{(}{)}{;}{a}{}{,}

\begin{document}

\title{Efficient photochemistry of coronene:water complexes.}

\authorrunning{Noble et al.~}
\titlerunning{Coronene:water photochemistry}

\author{J.~A. Noble$^a$, C. Jouvet$^b$, C. Aupetit$^a$, A. Moudens$^c$, and J. Mascetti$^a$}

 \institute{a Institut des Sciences Mol\'{e}culaires (ISM), Universit\'{e} de Bordeaux and CNRS, 351 Cours de la Lib\'{e}ration, F-33405 Talence, France.\\
   b CNRS, Aix-Marseille Universit\'{e}, UMR-7345, Physique des Interactions Ioniques et Mol{\'e}culaires (PIIM): 13397 Marseille Cedex 20, France.\\
   c LERMA-LAMAp, Universit\'{e} de Cergy-Pontoise, Observatoire de Paris, ENS, UPMC, UMR 8112 du CNRS, 5 mail Gay Lussac, 95000 Cergy Pontoise Cedex, France.\\
              \email{jennifer.noble@u-bordeaux.fr}
              }

   \date{Received \today; submitted }

\abstract
{The photochemistry of ices with polycyclic aromatic hydrocarbons (PAHs) has been extensively studied, but to date no investigation has been made of PAHs in interaction with low numbers (n $<$ 4) of molecules of water.
We performed photochemical matrix isolation studies of coronene:water complexes, probing the argon matrix with FTIR spectroscopy.
We find that coronene readily reacts with water upon irradiation with a mercury vapour lamp to produce oxygenated PAH photoproducts, and we postulate a reaction mechanism via a charge transfer Rydberg state.
This result suggests that oxygenated PAHs should be widely observed in regions of the ISM with sufficiently high water abundances, for example near the edges of molecular clouds where water molecules begin to form, but before icy layers are observed, that is at A$_V <$~3. In order to explain the low derived observational abundances of oxygenated PAHs, additional destruction routes must be invoked.}


\maketitle

\section{Introduction}

Polycyclic aromatic hydrocarbons (PAHs) are believed to represent a significant reservoir of interstellar elemental carbon (f$_c\sim$~3.5~$\times$ 10$^{-2}$, \citet{Tielens08}) and to be widely distributed throughout the interstellar medium (ISM) in both the gas and solid phases \citep{Salama08}. 
The carbonaceous component of interstellar dust is assumed to be formed in the envelopes around carbon stars by a series of reactions, nucleations and/or aggregations that lead from small carbon chains, via small and medium-sized PAHs, to larger PAHs, aggregates and nanoparticles \citep{Pascoli00,Contreras13}.
Analysis of the aromatic interstellar band (AIB) emission features in the mid infrared (MIR) supports the hypothesis that the interstellar PAH population is centred around molecules of the size C$_{50}$--C$_{100}$ \citep{Allamandola1989}. PAHs smaller than this will be destroyed by UV irradiation \citep{Madden2000,Gordon08} in regions of high photon flux. However, in molecular clouds, such molecules may survive in icy grain mantles where they are somewhat shielded from radiation. It is postulated that, in the spectra of YSOs, up to 9~\% of the unidentified absorption in the 5 -- 8 $\mu$m range can be attributed to neutral PAHs frozen out in dust grain mantles; these molecules account for up to 9~\% of the cosmic carbon budget \citep{Hardegree-Ullman2014}. 

It has long been established, thanks to extensive laboratory studies, that PAHs embedded in water ices undergo photochemical reactivity \citep[e.g.][]{Bernstein1999, Bernstein2002,Gudipati2003a,Gudipati2004}. Subjected to VUV radiation, isolated PAH molecules in cryogenic matrices form cations \citep[e.g.][]{MendozaGomez1995,Hudgins1995}. Such ionisation, by Lyman-$\alpha$ photons at 121~nm, is also seen for PAHs in water ice \citep{MendozaGomez1995,Gudipati2003a}, in addition to the formation of oxidation products such as alcohols and ketones \citep{Bouwman2011a,Bouwman2011b}. More recent studies have demonstrated the role of concentration in the reactivity of small PAHs \citep{Cuyulle2014,Cook2015}; low concentrations of PAHs in water ice favour the formation of oxygenated PAHs, rather than erosion processes that form small molecules such as CO, CO$_2$, and H$_2$CO \citep{Cook2015}. 

Water ice is believed to lower the barrier to photoionisation of PAHs \citep{Woon2004,Gudipati2004}, the recognised explanation for the formation of oxygenated photoproducts upon irradiation with a (low-energy) mercury lamp ($\lambda >$ 235~nm, \citet{Guennoun11a,Guennoun11b}). In this work, we test the hypothesis that an ice environment is necessary for the reaction of PAHs with water to occur upon irradiation by photons of energy $\lambda >$ 235~nm. We study the photochemistry of complexes of the PAH coronene (C$_{24}$H$_{12}$) with water by the matrix isolation technique. Complexes of one coronene molecule with up to four water molecules (C$_{24}$H$_{12}$:H$_2$O$_n$, n= 0 -- 4) were isolated in an argon matrix at 10~K and irradiated with a mercury lamp. The coronene molecule is often used in the laboratory as a model for larger, more interstellar-relevant PAHs \citep[e.g.][]{Joblin94,Oomens01,Guennoun11b}. Although smaller PAHs (n$_C <$ 50) are unlikely to survive the radiation conditions in the ISM, coronene is one of the largest, most UV-resistant species easily manipulated in the laboratory.
This is a standard experimental approach, due to the difficulties of studying larger PAH molecules in the laboratory. A mercury lamp, coupled with CsI windows in the experimental chamber, is used for irradiation in order to limit the energy of the incident photons to $\lambda >$ 235~nm (5~eV), thus ensuring that coronene can not be directly photoionised by the incident photons (ionisation potential 7~eV). We show, for the first time, that oxygenated photoproducts can form in the absence of a water ice, and propose a mechanism involving a charge transfer Rydberg state to explain this reactivity.

\section{Experimental methodology and theoretical calculations}\label{sec-expt}
Experiments were performed in a high vacuum experimental setup consisting of a stainless steel chamber (base pressure 10$^{-7}$~mbar) containing a CsBr substrate cooled to 10~K by a closed-cycle He cryostat. Infrared spectra were obtained in transmission mode using a Bruker 70 V Fourier Transform InfraRed (FTIR) spectrometer with a DTGS MIR detector. Spectra were recorded from 4000 -- 500~cm$^{-1}$ at a resolution of 0.5~cm$^{-1}$ with an average of approximately 100 scans. Argon was dosed from a gas line at a rate of 1~ml\,min$^{-1}$. Coronene (97~\%, Sigma-Aldrich) was co-deposited by sublimation of coronene powder heated in the oven to 180~$^o$C. The coronene had previously been heated to 120~$^o$C for several hours to remove excess water (the molecule is hygroscopic). Despite the heating, traces of H$_2$O remained in the coronene sample, so no additional H$_2$O was added to the mixing ramp before deposition, ensuring the lowest coronene:water ratio possible. Deposition lasted 155~minutes. The deposited C$_{24}$H$_{12}$:H$_2$O:Ar matrix was irradiated with a mercury lamp for a total of eight minutes (average power $\sim$~150~mW, total fluence was 0.72~J~m$^{-2}$), with IR spectra measured after one, three, and eight minutes of irradiation.

\textit{Ab initio} calculations were performed to determine the electronic structure of isolated coronene and of complexes of coronene with one, two or three water molecules (C$_{24}$H$_{12}$:H$_2$O$_n$; n= 0 -- 3). These calculations were performed with the TURBOMOLE programme package \citep{Ahlrichs89}, making use of the resolution-of-the-identity (RI) approximation for the evaluation of the electron-repulsion integrals \citep{Hattig03}. The equilibrium geometry of the ground states of the coronene molecule and of the coronene:water complexes were initially determined at the DFT level (B3lyp(+disp)/cc-pVDZ), then at the MP2 (M{\o}ller-Plesset second order perturbation theory) level. 
Adiabatic excitation energies of the lowest excited singlet states were determined at the RI-CC2 and/or RI-ADC(2)(second order Algebraic
 Diagrammatic Construction level \citep{Schirmer82}). Calculations were performed with the correlation-consistent polarised valence double-zeta (aug-cc-pVDZ) basis set \citep{Woon93} on H and O. 

\section{Results and discussion}\label{sec-results}

\begin{figure}[htb!]
\centering
\includegraphics[width=0.5\textwidth]{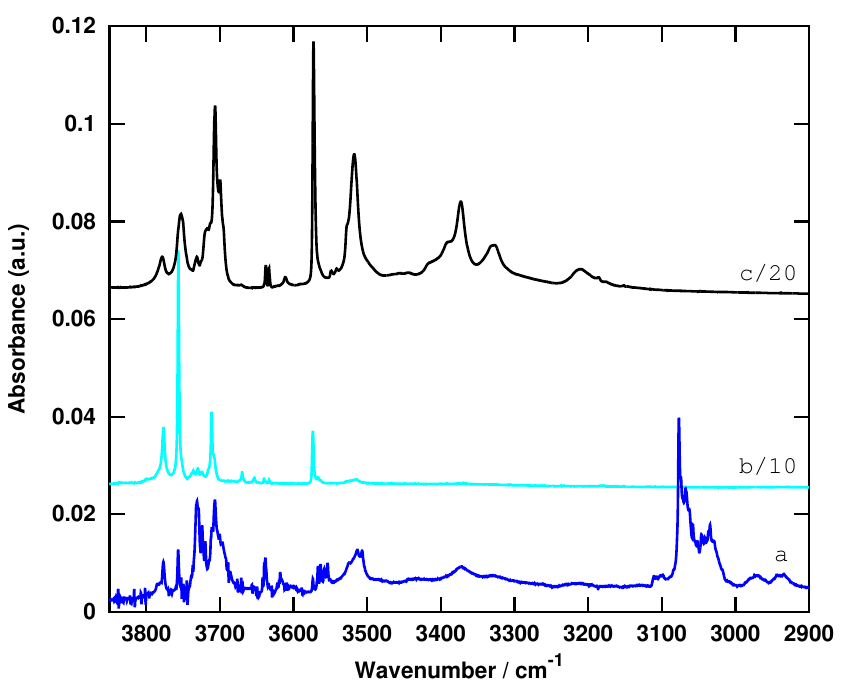}
\caption{FTIR spectra in the spectral range 3850 -- 2900~cm$^{-1}$ of (a) a coronene:water:argon matrix codeposited at 10~K, (b) a water:argon matrix of concentration $\sim$~1:500 deposited at 10~K and divided by a factor of ten, and (c) a water:argon matrix of concentration $\sim$~1:50 deposited at 10~K and divided by a factor of ten. The attributions of water and coronene for spectrum (a) are presented in Table~\ref{TableAttributions}.}
\label{fig0}
\end{figure}

Figure~\ref{fig0}a (blue) presents the spectrum of the co-deposited C$_{24}$H$_{12}$:H$_2$O:Ar matrix in the frequency range 3850 -- 2900~cm$^{-1}$. It is evident from the low absorbances of the bands of both H$_2$O and C$_{24}$H$_{12}$ that the matrix is very dilute. This is confirmed by comparison with the two reference spectra in Figure~\ref{fig0}: a H$_2$O:Ar matrix at a concentration of $\sim$~1:500 (cyan, trace b, 118 minutes deposition), and a H$_2$O:Ar matrix at a concentration of $\sim$~1:50 (black, trace c, 63 minutes deposition), which are both divided by a factor of ten in order to plot all three traces on the same absorbance scale.

A full list of spectral band attributions of the C$_{24}$H$_{12}$:H$_2$O:Ar matrix is provided in Table~\ref{TableAttributions}. The coronene is in the form of monomers and has not formed coronene clusters \citep{Roser15}. It shows characteristic absorption features in the 3100 -- 2900~cm$^{-1}$ region. In our C$_{24}$H$_{12}$:H$_2$O:Ar matrix we observe small shifts of coronene band positions, in particular those related to the edge of the molecule. For example, compared to literature values for coronene in argon matrices, the $\gamma_{CH}$ at 856~cm$^{-1}$ is red-shifted by one wavenumber, and the $\nu_{CC}$ band at 1319~cm$^{-1}$ is blue-shifted by two wavenumbers \citep{Vala93,Hudgins98}. By comparison with DFT calculations, as will be detailed in a forthcoming article (Noble et al. {\it in prep.}), we attribute these spectral signatures to coronene:water complexes isolated in the matrix.
Due to matrix effects, these shifts are much smaller than the effects of complexation predicted in the gas phase by DFT calculations \citep{Simon12}. Additionally, coronene:water complexes are limited to edge-on configurations by the argon matrix.

H$_2$O is present in monomeric and dimeric form, with additional bands corresponding to trimers and tetramers, its absorption features falling in the range between 3800 and 3300~cm$^{-1}$. There is no water ice present in the sample. The attributions of the absorption features are made by comparing the C$_{24}$H$_{12}$:H$_2$O:Ar matrix (Figure~\ref{fig0}, trace a) to the two reference depositions (traces b and c) and to previous matrix studies of H$_2$O \citep{Bentwood80,Perchard01,Coussan98}. The influence of trace amounts of N$_2$ in the matrix is more evident in the C$_{24}$H$_{12}$:H$_2$O:Ar matrix than in the H$_2$O:Ar matrices due to the very low relative concentration of the other molecular species, with absorption features attributed to (H$_2$O)$_{(1,2)}$:(N$_2$)$_{(1,2)}$ \citep{Coussan98}. There is clear evidence of complexation of water molecules in the C$_{24}$H$_{12}$:H$_2$O:Ar matrix. The spectrum much more closely resembles that of the higher concentration 1:50 H$_2$O:Ar matrix (trace c) than that of the 1:500 matrix (trace b), in terms of both spectral features present and the peak position of these bands, despite the fact that the water concentration in the C$_{24}$H$_{12}$:H$_2$O:Ar matrix is orders of magnitude lower, as can be determined by comparing the relative absorbances of all three matrices. It is not possible to directly calculate the water concentration in the C$_{24}$H$_{12}$:H$_2$O:Ar matrix, as the water is co-deposited with the coronene during heating of the oven, likely due to desorption of water molecules from the warm metal surfaces of the oven. 
However, a calibration experiment depositing coronene and water without argon was used to determine that the relative concentration of coronene to water in the deposited matrix is lower than 1:15.
It is possible that some of the co-deposited water is present in the form of water clusters isolated in the argon matrix, but we suggest that the spectral signatures of dimers, trimers and tetramers, more closely resembling a high concentration H$_2$O:Ar matrix deposition, is mostly due to complexation of water with coronene molecules. This also agrees with the shifts observed in the coronene absorption bands due to interaction with water.

\begin{figure}[htb!]
\centering
\includegraphics[width=0.5\textwidth,clip, trim=1.5cm 1.8cm 11.0cm 19.0cm]{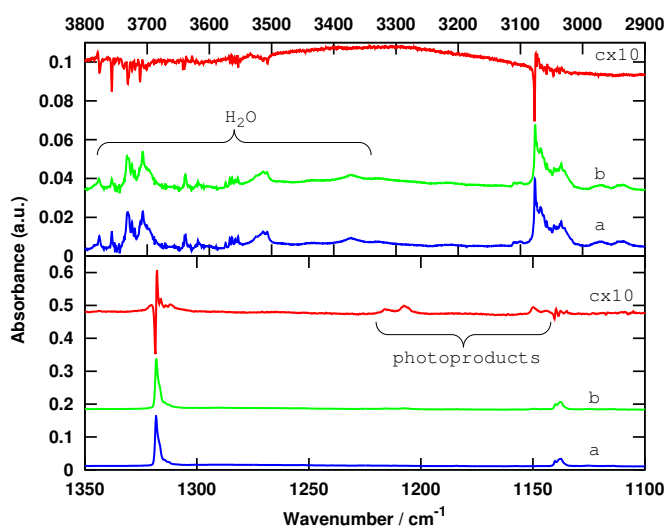}
\caption{FTIR spectra of a coronene:water:argon matrix in the spectral ranges 3800 -- 2900~cm$^{-1}$ (top panel) and 1350 -- 1100~cm$^{-1}$ (lower panel). The spectra are (a) codeposited coronene:water:argon matrix at 10~K, (b) the same matrix after three minutes of irradiation with a mercury lamp, (c) the difference spectrum (b)-(a), multiplied by ten. The attributions of water, coronene, and photoproducts are presented in Table~\ref{TableAttributions}, and the bands attributed to water and photoproducts are highlighted.}
\label{fig1}
\end{figure}

Figure~\ref{fig1}a (blue) presents the same spectrum of the initial C$_{24}$H$_{12}$:H$_2$O:Ar matrix in the frequency ranges 3800 -- 2900~cm$^{-1}$ and 1350 -- 1100~cm$^{-1}$. Figure~\ref{fig1}b (green) presents the spectrum of the C$_{24}$H$_{12}$:H$_2$O:Ar matrix after three minutes irradiation, and Figure~\ref{fig1}c (red) presents the difference spectrum (b-a) multiplied by a factor of ten. In this spectrum it is evident that there are various modifications of the spectral features: water bands have decreased slightly, the coronene bands have been perturbed and there are new bands present at 1215, 1207 and 1150~cm$^{-1}$. A previous study by \citet{Guennoun11b} determined that coronene reacts with water in an amorphous water ice after irradiation with a mercury lamp. The major photoproduct was determined to be dihydroxycoronene (C$_{24}$H$_{12}$O$_2$), produced due to reaction of water at the edges of the coronene molecule. The bands at 1215, 1207 and 1150~cm$^{-1}$, which are observed after three minutes irradiation of our coronene:water:argon matrix, are attributed to the same oxygenated photoproduct, dihydroxycoronene. The details of the attribution are presented in Table~\ref{TableAttributions}.

\begin{table}[htb!]
\small
  \caption{Attributions of matrix isolated molecular species observed in this study.}\label{TableAttributions}
  \begin{tabular*}{0.5\textwidth}{@{\extracolsep{\fill}}lll}
\hline
Molecular          & Observed band(s)          & Attribution\\
species             & cm$^{-1}$ & \\
\hline
H$_2$O             & 3777, 3757, 3724 & $\nu_3$\tablefootmark{a,b}\\
(H$_2$O)$_2$    & 3707, 3574              & $\nu_3$, $\nu_1$\tablefootmark{a,b}\\
(H$_2$O)$_3$    & 3525, 3513              & $\nu_1$\tablefootmark{a}\\ 
(H$_2$O)$_4$    & 3373                           & $\nu_1$\tablefootmark{a}\\ 
Coronene           & 3077, 3068, 3034 & $\nu_{CH}$\tablefootmark{c}\\
                          & 1319                           & $\nu_{CC}$\tablefootmark{c}\\
                          & 856                             & $\gamma_{CH}$\tablefootmark{c}\\
N$_2$ (trace)     & 3731, 3724, 3719       & (H$_2$O)$_{(1,2)}$:(N$_2$)$_{(1,2)}$\tablefootmark{d}\\
                         & 3639, 3567, 3554        & (H$_2$O)$_2$:N$_2$\tablefootmark{d}\\
\hline
Photoproduct 1 & 1405, 1383, 1215, 1207  & dihydroxycoronene \tablefootmark{e}\\
                         & 1150, 1029, 964                 & \\
Photoproduct 2 & 1579, 1221, 994    & 1,10-coroquinone \tablefootmark{e}\\
\hline
  \end{tabular*}

\tablefoottext{a}{\citet{Bentwood80}}
\tablefoottext{b}{\citet{Perchard01}}
\tablefoottext{c}{\citet{Hudgins98}}
\tablefoottext{d}{\citet{Coussan98}}
\tablefoottext{e}{\citet{Guennoun11b}}
\end{table}

\begin{figure}[htb!]
\centering
\includegraphics[width=0.5\textwidth]{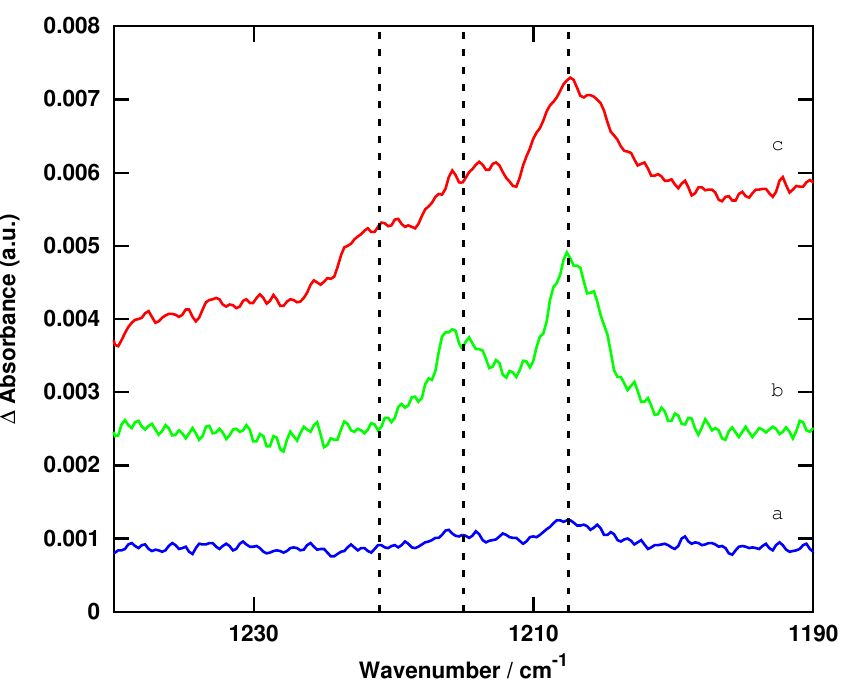}
\caption{FTIR difference spectra of an irradiated coronene:water:argon matrix. The spectra are presented as difference spectra with respect to the deposited matrix, and represent total irradiation times of (a) one minute, (b) three minutes, and (c) eight minutes. Photoproduct bands are observed at the frequencies 1207, 1215 and 1221~cm$^{-1}$ and are attributed to oxygenated PAHs, as described in the text and tabulated in Table~\ref{TableAttributions}.}
\label{fig2}
\end{figure}

Figure~\ref{fig2} shows the difference spectra of the C$_{24}$H$_{12}$:H$_2$O:Ar matrix after irradiation for one, three and eight minutes. There is already some evidence of photoproducts after only one minute of irradiation. The photoreactivity is a further proof which retroactively confirms the presence of coronene:water complexes in the deposited matrix. The absorption bands at 1207 and 1215~cm$^{-1}$ are present in spectra (a) and (b), while in spectrum (c) there is no further growth of these peaks, suggesting that the photoreaction is saturated after eight minutes irradiation. However, a third feature at 1221~cm$^{-1}$ appears. This seems to be the signature of another photoproduct which forms after the formation of dihydroxycoronene. In the previous study by \citet{Guennoun11b}, the formation of 1,10-coroquinone was observed after four hours of irradiation of a C$_{24}$H$_{12}$:H$_2$O ice mixture. Although the band observed here is redshifted by $\sim$~13~cm$^{-1}$ compared to the band attributed to 1,10-coroquinone in amorphous water ice, the presence of additional bands at 1579 and 994~cm$^{-1}$ could also be attributable to this molecule. We therefore tentatively attribute the band to the formation of the secondary photoproduct as 1,10-coroquinone, following the assignment by \citet{Guennoun11b}. It should be noted that a small amount of ice is observed in the IR spectrum after eight minutes of irradiation. It is unclear whether this is due solely to background deposition of trace amounts of H$_2$O on top of the matrix or due to some diffusion of water molecules within the matrix. By analysis of the small increase in absorbance centred at 3300~cm$^{-1}$, we can place a limit of 3~$\times$~10$^{17}$ molecules\,cm$^{-2}$ of H$_2$O deposited from the chamber between the initial matrix deposition and eight minutes irradiation, that is $\sim$~2.5 ML\,min$^{-1}$, which is attributed to a low level of residual outgassing of H$_2$O from heated metal surfaces in the vicinity of the PAH oven, and is reasonable for a high vacuum system with a base pressure of 10$^{-7}$~mbar at 10~K. This is equivalent to approximately one third the number of water molecules deposited in the same length of time during the coronene:water:argon matrix deposition.

These results are the first experimental evidence of the photoreactivity of small PAHs in complexes with water. Previous studies have concentrated on PAHs embedded in, or adsorbed to, water ices \citep[e.g.][]{Bernstein1999, Bernstein2002,Gudipati2003a,Gudipati2004, Guennoun11a, Cook2015}. Here we show that an ice matrix is not necessary to promote the photoreactivity of coronene with water, even using a ``soft'' UV source like a mercury lamp ($\lambda >$ 235~nm, average power $\sim$~150~mW).


Previous studies using a mercury lamp to irradiate a mixed (or layered) PAH:water ice have demonstrated that the production of oxygenated photoproducts occurs for energies lower than the ionisation energy of the PAH \citep{Gudipati2004,Guennoun11b}. In those works, the presence of an ice surface was believed to lower the barrier to the ionisation of the PAH molecule, which then reacts with water. In our study, no ice is present and yet we observe the formation of oxygenated photoproducts. The RI-CC2 calculations carried out on coronene:water complexes provide an alternative explanation for this reactivity in the form of a charge transfer (CT) Rydberg state.

\begin{table}[htb!]
\small
  \caption{Calculated energies (eV) for the $\pi\pi^*$ and $\pi\sigma^*$ excited states of C$_{24}$H$_{12}$:H$_2$O$_n$ (n= 0 -- 3) complexes in the Cs energy constraint.}\label{TableEnergies}
  \begin{tabular*}{0.5\textwidth}{@{\extracolsep{\fill}}lllll}
\hline
n & 0 & 1 & 2 & 3 \\
\hline
$\pi\pi^*$ vertical & 3.27 & 3.28 & 3.28 & 3.26\\
$\pi\sigma^*$ vertical & 5.27 & 5.09 & 5.19 & 4.78\\
$\pi\pi^*$/$\pi\sigma^*$ vertical gap & 2.00 & 1.81 & 1.91 & 1.52\\
Cs adiabatic gap & -- & 1.79 & 1.57 & 1.44\\
\hline
  \end{tabular*}
\end{table}

\begin{figure}[htb!]
\centering
\includegraphics[width=0.5\textwidth,clip, trim=3.5cm 0.5cm 3.5cm 2.0cm]{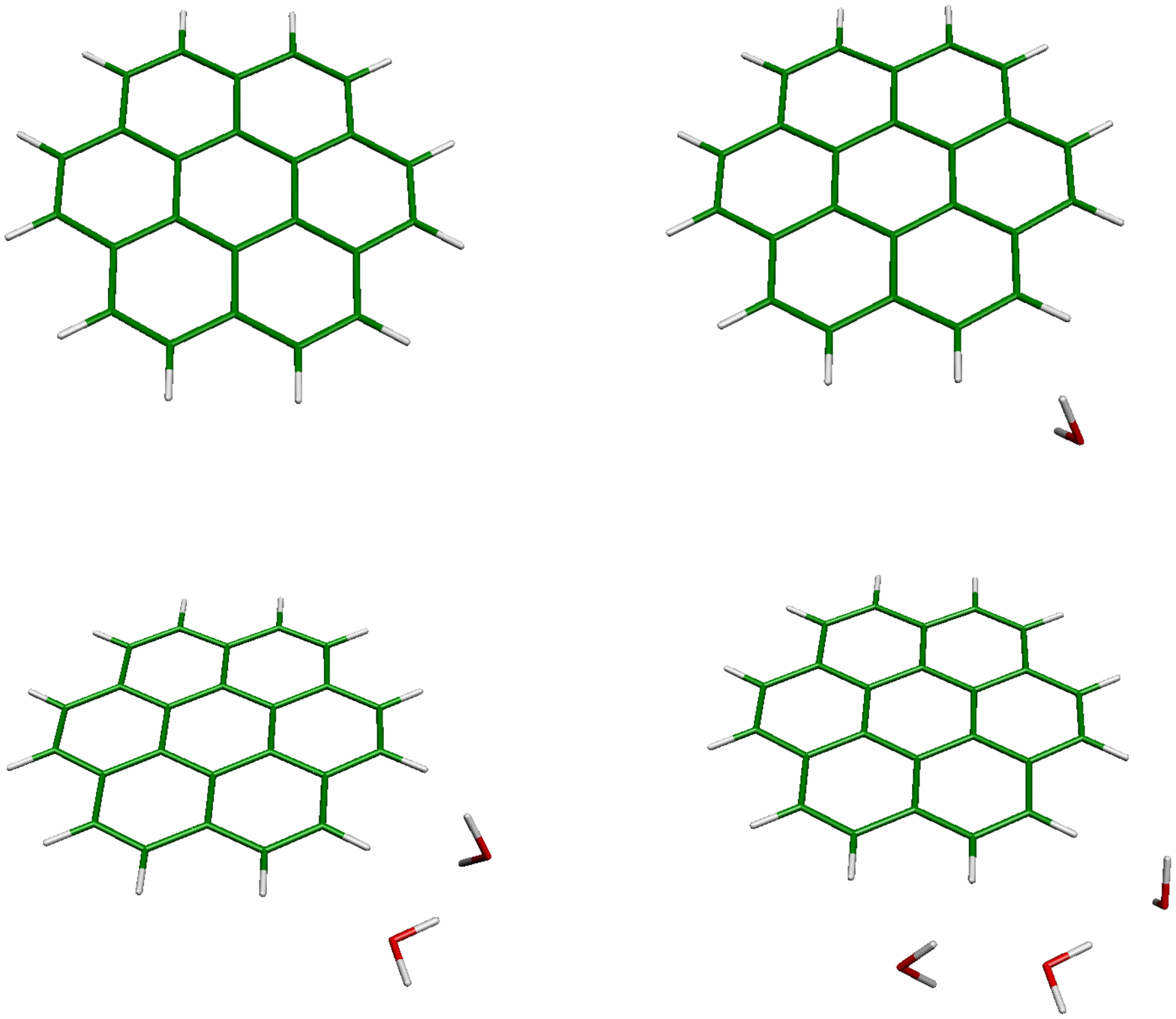}
\caption{The optimised geometries of the C$_{24}$H$_{12}$:(H$_2$O)$_n$ complexes, n = [0,3].}
\label{fig3}
\end{figure}

Table~\ref{TableEnergies} presents the calculated energies of the first excited states of coronene:water complexes. 
Figure~\ref{fig3} shows the structures and orbitals of the calculated complexes in different electronic states. In all calculations the geometry of the complexes is planar, that is, water is positioned at the edge of the coronene molecule and not, for example, above the plane of the aromatic system. This is an appropriate model for two reasons: firstly, reactivity between water and PAHs occurs at C-H groups around the edge of the PAH molecule, producing alcohols or ketones; and secondly, in our matrix experiments the position of water relative to coronene is constrained to edge-on configurations by the presence of the argon matrix (which will be described in detail in a forthcoming joint experimental-theoretical publication).

In substituted aromatic molecules (such as phenols or indoles), excited state reactivity is due to the presence of a CT state, which leads to H loss from the aromatic molecule. In a planar molecule (Cs symmetry) this CT state has a $\sigma$ symmetry and Rydberg-type character ($\pi\sigma^*$ state). In substituted aromatic solvents ({\it e.g.} NH$_3$ or H$_2$O), upon formation of complexes, the electron will be located in a very diffuse orbital on the solvent molecule. In the ground state equilibrium geometry, the state responsible for the reactivity is typically located 0.5 to 1~eV above the first excited S1 ($\pi\pi^*$) state. Based on this premise, we searched for the presence of such a CT Rydberg state in coronene:water complexes. In order to describe the diffuse orbitals involved, we chose to work in Cs symmetry, limiting the CT state of interest to one of the first a'' ($\pi\sigma^*$) states, independent of the number of the a' ($\pi\pi^*$) state below. We studied the evolution of the $\pi\pi^*$/$\pi\sigma^*$ energy gap as a function of complex size, retaining the Cs symmetry of the complex. The lower $\pi\pi^*$ state remains at approximately the same energy upon addition of water molecules to the complex ($\sim$~ 3.27~eV). The higher $\pi\sigma^*$ state, however, lowers in energy as the complex increases in size, from 5.27~ eV when n = 0 to 4.78~eV when n = 3. This is unsurprising, since CT states are strongly stabilised by polar molecules, while the stabilisation of the covalent excited state should be similar to that of the ground state. The adiabatic energy gaps were obtained by optimising the first excited state geometries in each symmetry and show an even clearer decrease with increasing complex size.

\begin{figure}[htb!]
\centering
\includegraphics[width=0.5\textwidth,clip, trim=2.2cm 20.5cm 22.05cm 3.2cm]{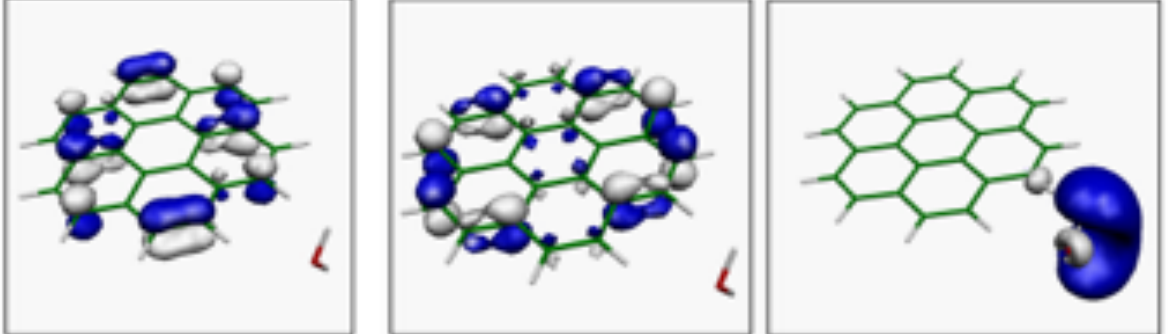}
\caption{The molecular orbitals calculated for the C$_{24}$H$_{12}$:H$_2$O complex in its (left to right) $\pi$, $\pi^*$, and $\sigma^*$ states.}
\label{fig4}
\end{figure}

Although there are limitations to these calculations in both size and symmetry, they offer an interesting explanation for the photochemical reactivity of coronene:water complexes. 
Excitation by an incident UV photon could promote an electron to the $\pi\sigma^*$ CT state. This could lead to the formation of the ion pair structure that may react, as previously postulated for coronene in ices \citep{Guennoun11b}, through the intermediate cation. If this were the mechanism, one would expect to observe the cationic intermediate in the experimental spectra, which was not the case in these data. Alternatively, the reaction could proceed via the H transfer mechanism, leading to a radical reaction; the coronene would lose a H and the H$_3$O intermediate would be formed, as follows: 

\begin{equation}
\begin{split}
C_{24}H_{12}(H_2O)_n + h\nu \rightarrow C_{24}H_{12}^\ast(H_2O)_n \rightarrow  \\
C_{24}H_{11}^{\ast\circ}H_3O^\circ(H_2O)_{n-1} \rightarrow C_{24}H_{11}OH(H_2O)_{n-1} + H_2. 
\end{split}
\end{equation}

Such an excited state reaction leading to the formation of the hydronium radical has recently been demonstrated in the aminophenol:water van der Waals complex \citep{Hernandez2015}. The planar geometry imposed during the calculation is likely to be a reasonable model for the structural constraints imposed upon coronene:water complexes formed in an argon matrix, where water molecules will favour edge-on positions (Noble et al. {\it in prep.})

In this study we have shown that PAH:H$_2$O complexes in argon matrices are photochemically active. In order to determine the extent to which a matrix effect may play a role in this reactivity, gas phase studies of PAH:H$_2$O complex photoreactivity are required. It is also important to study other PAH species to understand the effect of PAH size and/or geometry on their reactivity.

\section{Astrophysical implications}\label{sec-astro}

The major astrophysical implication of this study is that small PAH molecules can undergo photoreactivity with water molecules under the form of small complexes rather than, as was previously believed, uniquely in the presence of a water ice. This gives rise to the possibility of the formation of oxygenated PAH molecules in interstellar environments with low water abundances or higher temperatures (\textit{i.e.} where water does not condense onto grain surfaces in the form of an icy layer, but might be present in the form of a few water molecules.) An example of such an environment is PAHs or clusters of PAHs in the form of very small grains \citep{Rapacioli06} near the edge of a molecular cloud, where the interstellar radiation field is not completely attenuated, but molecules such as water begin to form on grain surfaces.

It is believed that small PAHs (n$_C <$ 50), such as our model PAH coronene, cannot survive the intense radiation fields of the ISM and are destroyed upon irradiation. However, larger PAH molecules are resistant to such radiation due to their capacity to redistribute energy in their various vibrational modes. We have shown that oxygenation of the C-H groups at the edge of PAHs is possible in coronene:water complexes, and it is conceivable that this would also be the case for larger PAHs, therefore we would expect to observe oxygenated PAHs in a range of irradiated regions, such as the edges of molecular clouds and possibly even photon dominated regions (PDRs, although ionisation and complex dissociation will dominate in such regions.) For small and medium PAHs, the partial radiation shielding offered by a molecular cloud could prevent dissociation, but might allow for the lower energy oxygenation reaction pathways at extinctions of A$_V <$~3 if the PAH molecule is associated with one or more water molecules (but is not yet embedded in a water ice.) 

Previous studies have constrained the abundance of OH-substituted PAHs observed in the ISM as being of the order of 0.002 \citep{Tielens08}, determined by analysis of the 2.77~$\mu$m band. The coronene molecule is a convenient laboratory model for larger PAHs stable in the ISM, but is unlikely to be present itself at high abundances. Very few observational studies have provided spectroscopic evidence for the absorption features of PAHs in molecular clouds \citep[e.g.][]{Chiar00}. However, we can compare the photoproduct bands observed in our experiments to the AIBs observed in the ISM with the additional caveats that these band positions are measured in matrices and thus their frequencies are blue-shifted compared to gas-phase values (by up to $\sim$~ 30~cm$^{-1}$ \citep{Joblin94}), that there is likely some attenuation of the vibrational modes by the presence of the cryogenic matrix \citep{Joblin94} and that there will likely be changes in band profiles, widths, frequencies (red-shifts of approximately 15~cm$^{-1}$) and relative intensities, due to the emission process, that differ from those in absorption spectra \citep[e.g.][]{Boersma13}. The highest intensity bands attributed to the oxygenated photoproduct dihydroxycoronene lie at 1207, 1215 and 1150~cm$^{-1}$. These bands fall within the spectral range corresponding to the 7.7 and 8.6~$\mu$m AIBs, and therefore this oxygenated PAH could potentially contribute to these features. However, the presence of other lower-intensity experimental bands, particularly that observed at 964 cm$^{-1}$, limits the potential interstellar abundance of oxygenated coronene, because this could only contribute in the far wing of the 11.2~$\mu$m AIB. It should also be noted that the limitation on the presence of oxygenated photoproducts of coronene does not preclude the presence of larger oxygenated PAHs, but merely limits the potential abundance of dihydroxycoronene. In general, oxygenated PAHs can be spectroscopically differentiated from PAHs via the presence of bands related to C-O bonds. These bands typically fall into the 1200 -- 1600 cm$^{-1}$ range \citep{Cook2015,Bouwman2011a,Guennoun11b} and therefore, if oxygenated PAHs are present, they will contribute to the 6.2, 7.7, and 8.6~$\mu$m bands \citep{Peeters2002}. AIB signatures decrease rapidly in intensity at the diffuse to dense ISM transition (e.g. at the edge of molecular clouds) as the PAHs agglomerate and/or freeze out onto dust grains. Therefore, interstellar signatures of oxygenated PAHs will be difficult to observe at extinctions where PAH:water clusters could be present and the ISRF is not completely attenuated (A$_V \sim$~2).

In this study, we have shown that PAHs are highly reactive with water, readily forming OH- (and possibly C=O-) substituted molecules. Based on the results of this exploratory study into the reactivity of a PAH with a very small number of water molecules, we would perhaps expect the abundance of oxygenated PAHs to be higher, particularly in denser regions. The presence of destruction routes for these molecules should be investigated in more detail to help to understand the low derived interstellar abundances of oxygenated PAHs.

\begin{acknowledgements}
This work has been funded by the French national program Physique Chimie du Milieu Interstellaire (PCMI) and the Agence Nationale de la Recherche (ANR) project ``PARCS'' ANR-13-BS08-0005. We acknowledge the use of the computing facility cluster GMPCS of the LUMAT federation (FR LUMAT 2764).
\end{acknowledgements}

{}

\end{document}